\documentclass{aa}  
\usepackage{graphicx}
\usepackage{txfonts}
\pdfoutput=1
\begin{document}

\title{Detection of spatially structured scattering polarization of Sr I 4607.3 \AA \ with the Fast Solar Polarimeter}
\titlerunning{Spatially structured scattering polarization of Sr I 4607.3 \AA}

   \author{F. Zeuner
          \inst{1,2}
          \and
          A. Feller\inst{1}
           \and
          F. A. Iglesias\inst{1}
          \and
          S. K. Solanki\inst{1,3}
          }

   \institute{$^1$ Max-Planck-Institut für Sonnensystemforschung, 37077 Göttingen, Germany\\
   $^2$ Georg-August-Universität Göttingen, Friedrich-Hund-Platz 1, 37077 Göttingen, Germany\\
   $^3$ School of Space Research, Kyung Hee University, Yongin, Gyeonggi-Do, 446-701, Republic of Korea}

   \date{Accepted August 3rd, 2018}

 \abstract 
    {Scattering polarization in the Sr I 4607.3 \AA \ line observed with high resolution is an important diagnostic of the Sun's atmosphere and magnetism at small spatial scales. Investigating the scattering polarization altered by the Hanle effect is key to constraining the role of small-scale magnetic activity in solar atmospheric activity and energy balance. At present, spatially resolved observations of this diagnostic are rare and have not been reported as close to the disk center as for $\mu=0.6$.}
    {Our aim is to measure the scattering polarization in the Sr I line at $\mu=0.6$ and to identify the spatial fluctuations with a statistical approach.}
    {Using the Fast Solar Polarimeter (FSP) mounted on the TESOS filtergraph at the German Vacuum Tower Telescope (VTT) in Tenerife, Spain, we measured both the spatially resolved full Stokes parameters of the Sr I line at $\mu=0.6$ and the center-to-limb variation of the spatially averaged Stokes parameters.}
    {We find that the center-to-limb variation of the scattering polarization in the Sr I line measured with FSP is consistent with previous measurements. A statistical analysis of Stokes $Q/I$ (i.e., the linear polarization component parallel to the solar limb), sampled with 0.16\arcsec \ pixel$^{-1}$ in the line core of Sr I reveals that the signal strength is inversely correlated with the intensity in the continuum. We find stronger linear polarimetric signals corresponding to dark areas (intergranular lanes) in the Stokes $I$ continuum image. In contrast, independent measurements at $\mu=0.3$ show a positive correlation of $Q$/$I$ with respect to the continuum intensity. We estimate that the patch diameter responsible for the excess $Q/I$ signal is on the order of 0.5\arcsec-1\arcsec.}
    {The presented observations and the statistical analysis of $Q$/$I$ signals at $\mu=0.6$ complement reported scattering polarization observations as well as simulations. The FSP has proven to be a suitable instrument to measure spatially resolved scattering polarization signals. In the future, a systematic center-to-limb series of observations with subgranular spatial resolution and increased polarimetric sensitivity (<$10^{-3}$) compared to that in the present study is needed in order to investigate the change in trend with $\mu$ that the comparison of our results with the literature suggests.}


   \maketitle
%
\section{Introduction}

The quiet solar photosphere is thought to be filled by small-scale and weak magnetic fields, as predicted by small-scale dynamo simulations of \citet{Vogler2007} and \citet{Rempel2014}. To test this model, detailed measurements of the often weak magnetic field and its spatial distribution is required. However, observations draw an inhomogeneous picture on magnetic field strength, inclination and distribution, see \citet{Lagg2016a}, \citet{Stenflo2011} and reviews on this topic by \citet{deWijn2009}, \citet{Steiner2012} and \citet{Borrero2015}. Observationally determining properties of fields weaker than a few hundred Gauss, which are tangled at spatial scales close to the observations' spatial resolution is quite a challenge, as the Zeeman effect as a diagnostic tool has two major intrinsic disadvantages for extracting information about these fields. Firstly, the sensitivity to weak magnetic fields is low, particularly to the component perpendicular to the line-of-sight, and secondly, if the magnetic field is turbulent, signal cancellation within a resolution element is possible. This makes it particularly difficult to probe the magnetic field's properties in inter-network regions, as the results conspicuously depend on the method and spectral line employed, see e.g., \citet{Borrero2015}. To a large extent, these disadvantages are overcome by employing the Hanle effect, see for example, \citet{LandiDeglInnocenti} and \citet{Stenflo1994}. \par
A photospheric line widely used for Hanle diagnostics is Sr I 4607.3 \AA, which has the advantage of providing scattering signals at $\mu=0.1$ above 1\% (see, e.g., \citealp{Stenflo1997a} and \citealp{Gandorfer2002}). Here $\mu= \cos(\theta)$, where $\theta$ is the heliocentric angle and $\mu=0$ at the solar limb. So far, measurements with the required polarimetric sensitivity of the scattering polarization signals in this line suffer from insufficient spatio-temporal resolution (see \citealp{TrujilloBueno2007} and references therein). Additionally, the observed target needs to be sufficiently far from the solar limb to confidently correlate the granulation pattern with the detected polarimetric signals for interpretation. 
Observations with a spatial resolution of about 0.6\arcsec \ by \citet{Malherbe2007} at $\mu=0.3$ from the west limb display a positive correlation between the Stokes $Q$/$I$ signal in the Sr I line core and continuum intensity. Their findings possibly hint at a Hanle effect acting in the intergranular regions where higher magnetic fields are expected, in agreement with simulations carried out by \citet{Bueno2004}. However, close to the disk center,  \citet{TrujilloBueno2007} predicted theoretically that most of the scattering polarization is produced by local symmetry breaking of the radiation field by atmospheric inhomogeneities. In this case, the theoretically expected $Q$/$I$ signal fluctuations lie between $-0.08$\% and 0.9\%, as they would be observed with a 1 m telescope and a spectral resolution of 25 m\AA \ at $\mu=0.5$, without considering a depolarizing magnetic field. \par 
Here we present the results of two sets of observations carried out at the German Vacuum Tower Telescope (Tenerife, Spain) in the Sr I 4607.3 \AA \ line taken with the Fast Solar Polarimeter \citep{Iglesias2016} attached to the TESOS filtergraph \citep{Kentischer1998,Tritschler2002}. First we check if the data obtained with the Fast Solar Polarimeter are consistent with those in the literature by measuring the center-to-limb variation of the linear polarization in Sr I and in a neighboring Fe I line that does not show scattering polarization. This center-to-limb variation of the $Q$/$I$ amplitude is compared with published results. In a second step, we use a statistical approach to analyze photon-noise limited spatial fluctuations of the $Q$/$I$ signal at subgranular resolution and (1) search for a correlation between the $Q$/$I$ amplitude in the line core and continuum intensity and (2) estimate the structure size of scattering polarization signals.

\section{Observations and data reduction}

Our observations were obtained with a prototype of the Fast Solar Polarimeter (FSP) mounted at the Vacuum Tower Telescope (VTT) at the El Teide observatory on Tenerife, using the TESOS Fabry-Perot tunable filtergraph. This particular prototype of FSP is composed of a fast, cooled pnCCD camera synchronized with a  ferro-electric liquid crystal based polarization modulator. The image dimensions are 248$\times$256 pixel$^2$. The sampling of 0.08\arcsec pixel$^{-1}$ corresponds to approximately critical sampling at the diffraction limit of the 0.68 m VTT aperture. The TESOS bandpass has a FWHM of about 25 m\AA \ \citep{Beck2010b}. FSP is run with a polarization modulation frequency of 100 Hz, which corresponds to a frame rate of 400 frames s$^{-1}$ (four modulation states per cycle). For more details of the FSP and the used modulation scheme see \citet{Iglesias2016}. Calibration data, that is, dark images, a modulated flat field for all wavelength positions (moving the telescope randomly around disk center) and polarimetric calibration images were recorded within a few minutes of the observations. 

\subsection{Center-to-limb variation of scattering polarization}

\subsubsection{Observations}

On 6 May 2015 we observed quiet Sun regions at different limb distances between 17 and 19 UTC without using the adaptive optics (AO) system.  We recorded the full Stokes vector at 11 wavelength positions, whereby the full record of all positions is called a scan:  [$-90$, 90, $-60$, 60, $-40$, 40, $-20$, 20, $-10$, 10, 0] m\AA \ around a central wavelength. As central wavelengths we used $\lambda_{c,Sr}=4607.33$  \AA \ centered on the core of  the Sr I line and $\lambda_{c,Fe}=4607.63$  \AA \ centered on the core of a neighboring Fe I line. Ten scans through the Sr I line were made, each at a different solar disk position, ranging between solar disk center and the solar north limb in steps of $\Delta \mu = 0.1$. In this way we do not need to correct for Doppler shifts due to solar rotation. The Fe I line was scanned at $\mu=[1, 0.8, 0.27, 0.1]$. At  $\mu=0.1$ the solar limb is visible as a strong intensity drop in the upper part of the images. The individual scans took between 99 s and 106 s to complete and will be referred to as scan data hereinafter.

\subsubsection{Data reduction}
\label{sec:dare_clv}

For a detailed description of the basic reduction steps, see \citet{Iglesias2016}. 
The basic data reduction consists of three steps. They correct for offsets by subtracting a low-noise dark frame, image smearing, and common mode. The common mode signal is estimated from shielded pixels and subtracted from each semi-row. The image smearing, caused by shutter-free operation of frame-transfer CCD detectors, is corrected by using a modified standard model developed and implemented by \citet{Iglesias2015}. After the basic reduction, 800 frames within each modulation state and wavelength position are averaged to a single image. The flat field and the scan data are corrected and averaged identically. \par
For polarimetric calibration, FSP recorded 19 predetermined polarization states, generated in front of the modulator. Therefore, instrumental polarization caused by optical elements of the telescope located in front of the modulator is excluded from the calibration. The measured intensities were used to fit the elements of the over-determined $4 \times 4$ (de-)modulation matrix. The demodulation matrix was then applied to the flat field and the scan data in the same way. The polarimetric calibration is described in detail by \citet{Iglesias2016}. \par
The flat field and scan Stokes $Q$, $U$, $V$ images were normalized to Stokes $I$. The flat field $Q$/$I$, $U$/$I$ and $V$/$I$ images were then subtracted from the respective scan data.
This procedure was chosen to take into account artificial offsets in $Q$/$I$, $U$/$I$ and $V$/$I$ due to telescope polarization and to remove polarized fringes. As only very low polarization is expected in the continuum  \citep{Gandorfer2002}, we calculated spatial averages of the $Q$/$I$, $U$/$I$ and $V$/$I$ images taken at $-90$ m\AA \ and subtracted them as an offset from all wavelength points and corresponding polarization state images. A heuristic cross-talk removal was applied to correct the scan data for the uncalibrated telescope polarization, since there is no reliable telescope model available for the used wavelength range. We corrected for cross-talk between $V$/$I$ and $Q$/$I$, as well as between $V$/$I$ and $U$/$I$. This was done by rotating the whole FOV around the $U$ and $Q$ axes of the Poincaré sphere until the $V$/$I$ signal was minimized. The estimated cross-talk with this method is (9$\pm 2$)\% for $V$/$I$ to $Q$/$I$, corrected by a 5$^{\circ}$ rotation around the $U$ axis. For the cross-talk from $V$/$I$ to $U$/$I$ we estimated (57$\pm 1$)\%, corrected with a 35$^{\circ}$ rotation around the $Q$ axis. The direction of +$Q$/$I$ is parallel to the north solar limb. \par
We then averaged all pixels for a given $\mu$ value and for one wavelength. 
 Given that $\mu$ changes more rapidly over a given distance on an image with smaller $\mu$-values, the image closest to the limb was divided into equally sized stripes of width 1.2\arcsec that are parallel to the solar limb and the linear polarization within these stripes was averaged. The need for such narrow stripes is heightened by the fact that the $Q$/$I$ amplitude also increases most rapidly close to the limb. The solar limb was defined as the part of the image where the slope of the intensity was steepest. Since the AO system was off, the limb position changed in the field of view per wavelength position and was therefore obtained for each position separately.

\subsection{Spatially resolved Stokes measurements}

\subsubsection{Observations}

Our spatially resolved Stokes measurements in the Sr I line were performed on May 27 2014 between 11 and 13 UTC. The AO was locked on a quiet Sun region at $\mu=0.6$
toward the north solar limb. Our FOV is 20\arcsec \ by 20\arcsec. Five wavelength positions relative to 4607.3  \AA \ ( [$-90$, $-60$, 30, $-30$, 0] m\AA) \ were scanned. The exposure time was 1.25 s per modulation state and wavelength position for a single such scan, resulting in a total exposure time of 5 s per wavelength position. A single scan over all wavelengths required 32 s, implying a duty cycle of about 78\%. In total, ten such scans were obtained.

\subsubsection{Data reduction}
\label{sec:dare_spr}

 The basic data reduction, the demodulation and normalization of the Stokes images were done as described in Sect. \ref{sec:dare_clv}, including the flat-fielding procedure. Because of this procedure, which requires demodulation before flat-fielding, we were not able to make use of image restoration techniques to further increase the spatial resolution in the polarization Stokes images. From the demodulation matrix, we obtain 95\% total polarimetric efficiency \citep{DelToroIniesta2004}. For each scan all frames at a given wavelength position and modulation state were averaged.
 We determined the orientation of our Stokes reference system with respect to the solar limb by using a dataset acquired at 13 UTC in the Fe I 4607.6 \AA \ line, where we observed a pore region. 
 We rotated our pore images by $10^\circ$  anti-clockwise until they matched SDO/HMI Stokes images of the same region in both orientation and polarization, which for HMI are calibrated to have Stokes +$Q$/$I$ parallel to the north limb \citep{Schou2012}. 
   
\begin{figure}
                \centering
                \includegraphics[width=\hsize]{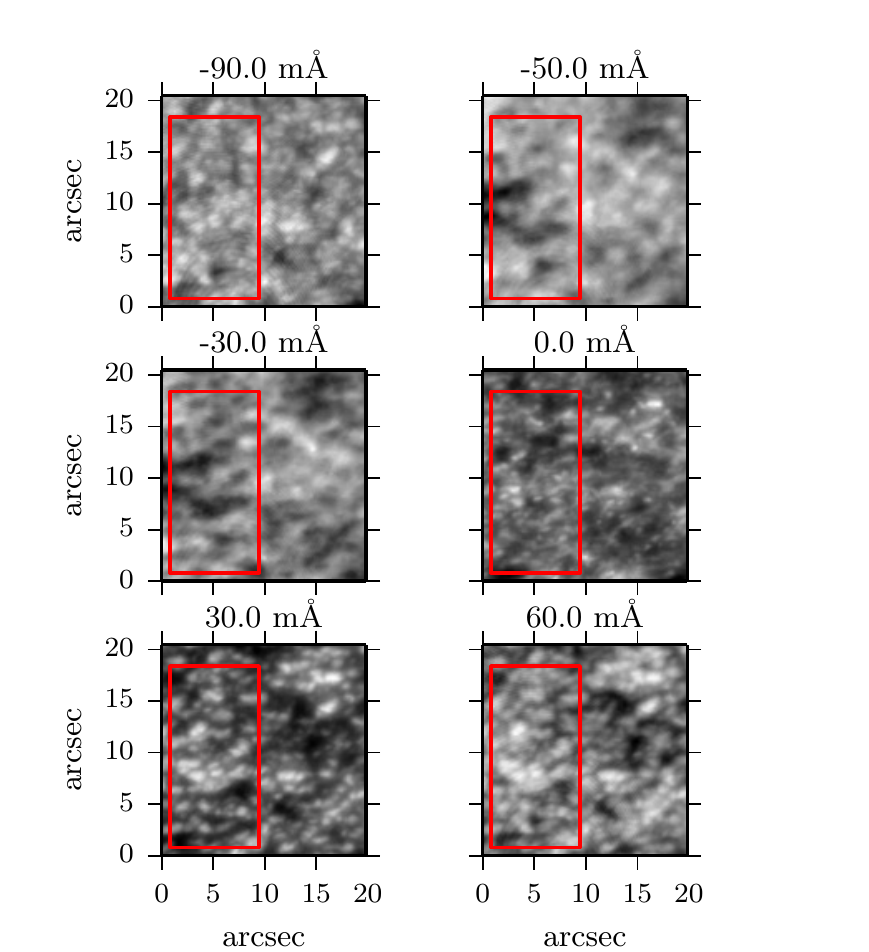}
                   \caption{
                   	Stokes $I$ images for different wavelengths at $\mu=0.6$. The region of interest for the analysis lies inside the red box, which excludes Zeeman signatures found in the right corner. The north solar limb is parallel to the shorter axis of the red box. The image labeled with 30 m\AA \ corresponds to the Sr I line core.
                           }
                      \label{fig:images}
                \end{figure}

We heuristically corrected for cross-talk between $V$/$I$ and $Q$/$I$, and $V$/$I$ and $U$/$I$ by minimizing a small longitudinal Zeeman signature at the edge of our FOV in $U$/$I$ and $Q$/$I$, using a Poincaré rotation, as in Sect. \ref{sec:dare_clv}. The used angles were 17$^{\circ}$ and 41$^{\circ}$ for rotation around the $Q$ axis and the $U$ axis, respectively, corresponding to cross-talk values of (29$\pm 2$)\% between $Q$/$I$ and $V$/$I$ and (75$\pm 1$)\% between $U$/$I$ and $V$/$I$. For further analysis of the scattering polarization, regions with a Zeeman signature were excluded (see  Fig. \ref{fig:images}), which resulted in a region of interest of 8\arcsec \ by 17\arcsec. Our data suffered from variable seeing conditions. For further analysis we therefore selected the first two scans with stable AO locking. To increase the signal-to-noise ratio in the polarization signal while not degrading spatial resolution significantly, we averaged both scans and applied a 2 $\times$ 2 spatial binning. Consequently, in the following we analyze one image per Stokes parameter and per wavelength position, where each image has a sampling of 0.16\arcsec \ pixel$^{-1}$ and 2.5 s exposure time.

\section{Results}

\subsection{Center-to-limb variation of scattering polarization}

\begin{figure}
   \centering
   \includegraphics[width=\hsize]{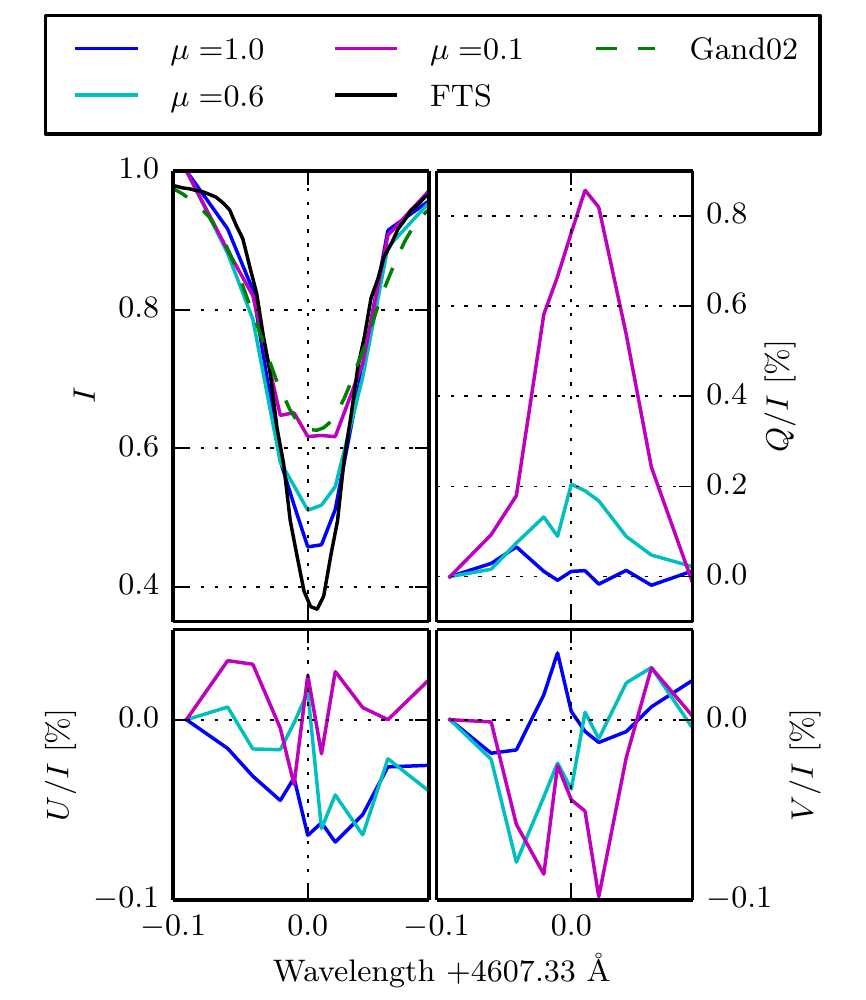}
      \caption{
      	Center-to-limb variation of the polarized profiles of the Sr I 4607.3 \AA \  line, spatially averaged over one image per $\mu$-value. To avoid cluttering, only three out of the ten measured $\mu$ positions are displayed. For comparison, the Stokes $I$ profile given in the FTS atlas (black solid line) and Stokes $I_{Gan02}$ taken from \citet{Gandorfer2002} (green dashed line) is included. Each Stokes $I$ spectrum is normalized to the continuum point at -90 m\AA. Residual polarimetric offsets after flat-fielding are removed by subtracting the value of  $Q$/$I$, $U$/$I$ and $V$/$I$ at -90 m\AA \ from the respective profiles at each wavelength position. 
              }
         \label{fig:linesr06}
   \end{figure}

Spatially averaged Sr I Stokes profiles for $\mu =$ 1.0, 0.6, 0.1 are plotted in Fig. \ref{fig:linesr06}. Each data point in the polarimetric spectral profiles has a low photon noise level (i.e., $\sigma I$ / <$I$> < 0.02\%) due to the spatial averaging. For Stokes $I$, we additionally plot Fourier Transform Spectrograph (FTS) atlas data by \citet{Neckel1999} and the Stokes $I_{Gan02}$ close to the limb at $\mu=0.1$ based on the atlas of \citet{Gandorfer2002}, which has been provided in electronic form by IRSOL as a compilation by \citet{fts}. The line depth of the Stokes $I_{Gan02}$ and the limb measurement are comparable, but for the FTS and our measured line at disk center there is a line depth difference of about 10\%. This is likely due to spectral scattered light as well as lower spectral resolution of TESOS compared to FTS. While Stokes $U$/$I$ and $V$/$I$ show no systematic correlation with limb distance within the spectral width of the scan in Fig. \ref{fig:linesr06}, the Stokes $Q$/$I$ signal systematically increases with lower $\mu$ values, as expected.

\begin{figure}
 
   \centering
   \includegraphics[width=\hsize]{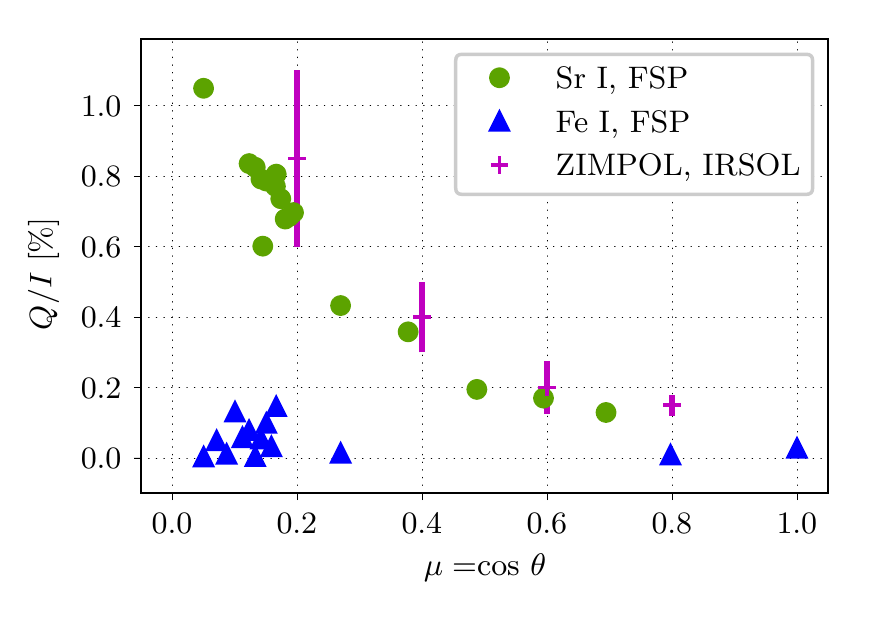}
      \caption{
      	Stokes $Q$/$I$ amplitude as a function of limb distance $\mu = \cos(\theta)$. The Sr I data points represent the amplitude of a Gaussian fit to the $Q$/$I$ line, while the Fe I data points are spectrally and spatially averaged values. We note that the scale is linear and that below $\mu=0.2$ less pixels were averaged for each data point. The signal-to-noise ratio beyond $\mu=0.7$ is below two and therefore the fitted amplitudes are discarded. For comparison, a range of Sr I polarization values obtained by ZIMPOL and a stationary IRSOL polarimeter (magenta bars show the range of values) are taken from \citet{Stenflo1997a}.
              }
         \label{fig:clv06}
   \end{figure}

 We fitted a Gaussian function to the $Q$/$I$ spectra for each $\mu$ position separately, and plot amplitudes with respect to $\mu$ in  Fig. \ref{fig:clv06}. For optimum fitting results the free parameters were not constrained. The standard deviation of 0.03\% for $Q$/$I$, $U$/$I$ and $V$/$I$ in the spectral profile at $\mu=1$ is taken to be an estimation of the noise level as no signal is expected. Amplitudes were discarded if their ratio with the aforementioned noise level was smaller than two. The low contrast of the intensity images (< 5\%) justifies the use of a constant weighting factor corresponding to the mean photon noise. The 95\% confidence interval for the amplitude is 0.2\%. We find a maximum $Q$/$I$ amplitude at $\mu=0.05$ of more than 1\%, which is in agreement with observations summarized by \cite{Malherbe2007}. Values in the literature of the $Q$/$I$ amplitude at this $\mu$ value range from 0.95\% \citep{Faurobert2001} to 1.5\% \citep {Stenflo1997}, which encompass our measured $Q$/$I$ amplitude. In  Fig. \ref{fig:clv06} we compare our $Q$/$I$ amplitudes for $\mu=0.2$, $0.4$ and $0.8$ with a range of values obtained by \citet{Stenflo1997a} with ZIMPOL and a stationary polarimeter installed at IRSOL. Our amplitudes at these $\mu$ values are again consistent with the measurements in the literature. Additionally, in Fig. \ref{fig:clv06}, we present as a blind test the $Q$/$I$ polarization we found in the neighboring Fe I 4607.6 \AA, averaged over the spectral width of the measurement, because the Stokes $Q$/$I$ is flat. As Fe I is insensitive to scattering polarization, the polarization signals do not depend on the limb distance. For $\mu$ < 0.2 we spatially averaged across a smaller spatial region. This results in a higher noise level, which explains the higher scattering of the Fe I data points below $\mu =0.2$.

 \subsection{Spatially resolved Stokes measurements}
 
 The Stokes $I$ images at each recorded wavelength position in the quiet Sun observation at $\mu=0.6$ are displayed in Fig. \ref{fig:images}. Based on the continuum Stokes $I$ images at -90 m\AA, we obtained a granulation intensity contrast of 2.5\%. To obtain the noise level for Stokes $Q$/$I$ per pixel the spatial power spectrum of the associated image is calculated. Based on the diffraction limit of the telescope, the cut-off frequency was computed. The power contained at spatial frequencies beyond the cut-off was assumed to be representative for the photon noise level. We obtained a noise level of 0.3\% per pixel for Stokes $Q$/$I$, which is consistent with the photon noise level estimated from Stokes $I$ images. We note that these images were averaged over two scans and spatially binned as described in Sect. \ref{sec:dare_spr}. At this noise level of 0.3\%, we do not find any direct visual evidence for spatial signal fluctuations at sub-granular scales in any polarization state.
 
 The amount of the line core polarization signal in Stokes $Q$/$I$ is close to 0\% in the granules, while for intergranules the values are around 0.25\%. The spatial mean of Stokes $Q$/$I$ in the line core is 0.1\%, which is lower than what is shown in Fig. \ref{fig:linesr06} for $\mu=0.6$ but close to the lower edge of the values found by \citet{Stenflo1997a}. The origin of this discrepancy is unknown, but possible spatial fluctuations should be unaffected by a shift of the mean signal.

  \subsubsection{Correlations with the continuum intensity}
    
\begin{figure}
      \centering
      \includegraphics[width=\hsize]{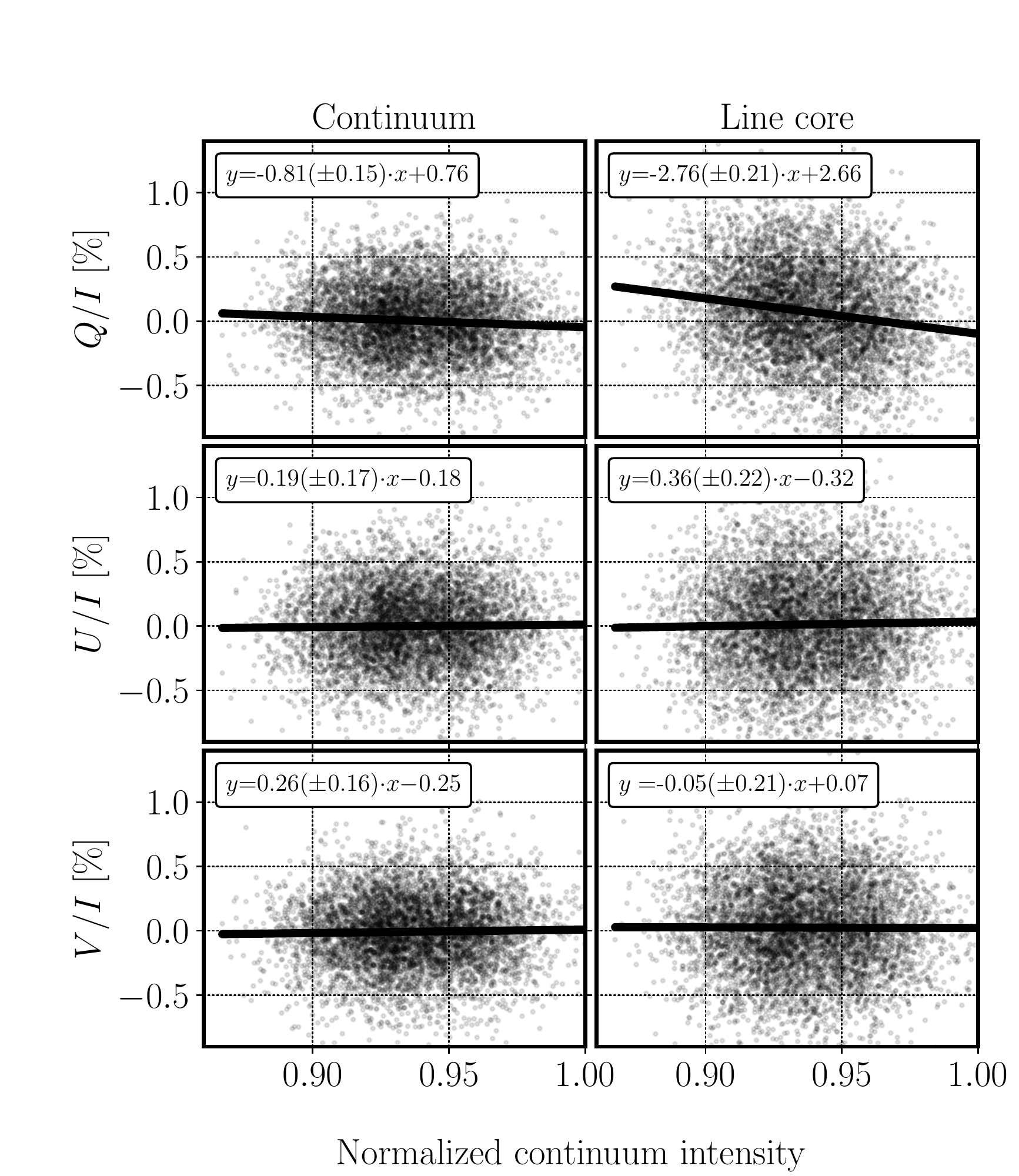}
         \caption{
         	Polarization signals of Stokes $Q$/$I$, $U$/$I$ and $V$/$I$ in the Sr I 4607.3 \AA \ line versus normalized Stokes $I$ continuum intensity. In the left panels the polarization found in the continuum is shown, while in the right panels the polarization signals from the line core are reported. Inserts display the function describing the linear fit, with $x$ as the normalized continuum intensity and $y$ the polarization percentage.
         }
            \label{fig:scatter}
      \end{figure}

 To reveal correlations of polarimetric signals in the line core with the continuum image intensity, scatter plots are shown in Fig. \ref{fig:scatter}. We find evidence that the Stokes $Q$/$I$ image in the Sr I line core is anti-correlated with the continuum intensity. To begin with, the correlation is quantified by fitting a linear function to the signals, estimating the slope and its standard error (67\% confidence interval). The values are given at the top of each panel in Fig. \ref{fig:scatter}. The hint of a correlation in $Q$/$I$ arises from the comparison of the $Q$/$I$ correlation in the line core with the correlations in the continuum (shown in the left panels of Fig. \ref{fig:scatter}) and with $U$/$I$ and $V$/$I$. The latter serve as references, as only photon noise is present in these images (see Sect. \ref{sec:ct_stokesi}). With $-2.76$ the slope of $Q$/$I$ in the line core exceeds more than 18$\times$ its standard error and is higher than the slope in any other polarization state and in the continuum. Higher noise in the line core due to fewer photons than in the continuum is represented by the larger error of the slopes. The weights for the linear fit are linked to the photon noise and assumed to be constant for each data point over the field of view, since the intensity contrast is low. \par
 To show the robustness of the slope against selection bias, we created random subsets with a quarter of the original sample size. The error in the slopes are then 2$\times$ larger than in the full data set. The slopes of the full data set are consistent with the slopes of the subsets within the error margins. We calculated the correlation coefficient for the $Q$/$I$ case to be $r=-0.17$.  The correlation coefficients for the reference images are at least one order of magnitude smaller, see Table \ref{tab:1} for a summary. A second statistical test was carried out by calculating the probability for observing a correlation coefficient as large or larger than $|r|$ under the (null) hypothesis that the parent population is uncorrelated. We follow the method in \citet{Bevington2003} where this two-sided probability $p$ for $N$ observations is given by

\begin{equation}
\label{eq:1}
p(r,N) = 2\int\limits_{|r|}^{1}{P(x,N) dx}.
\end{equation}

We approximated the normalized probability density function $P(r,N)$ for a correlation coefficient $r$ for large $N$ (i.e., $N>300$) with
\begin{equation}
\label{eq:2}
P(r,N) = \frac{(1-r^2)^{(N-4)/2}}{\int_{-1}^{1}(1-r'^2)^{(N-4)/2}dr'}.
\end{equation} 

 The calculated $p$-values for all polarimetric images are given in Table \ref{tab:1}.
 The null hypothesis is rejected if the $p$-value is smaller than $0.001$, which corresponds to a confidence level of 99.9\%. We reject the null hypothesis of no correlation for Stokes $Q$/$I$ in the Sr I line core, as the two-sided $p$-value is negligible. A small correlation is present for the $Q$/$I$ continuum image, potentially due to insufficient distance to the Sr I line core. However, the condition $p$>0.001 is fulfilled for the reference images. This situation is maintained when the above mentioned random subsets are analyzed. We are therefore confident that the anti-correlation with  $r=-0.17$ is statistically significant.\\

       \begin{table}
       	
       \caption{\label{tab:1} Pearson's correlation coefficients $r$ and the probability $p$ to observe a value as large or larger than $|r|$ for an uncorrelated population with $N$=6105.}
       \centering
       \begin{tabular}{l|c|c|c|c}
       \hline\hline
       Stokes & \multicolumn{2}{c|}{continuum} &  \multicolumn{2}{c}{line core} \rule{0pt}{2.6ex} \rule[-1.2ex]{0pt}{0pt} \\
       \hline
       & $r$ & $p$ & $r$ & $p$  \\ 
       \hline
      
        $Q$/$I$&$ -0.07$& $< 10^{-6} $&   $ -0.17$& $< 10^{-6}$  \rule{0pt}{2.6ex} \\ 
        $U$/$I$&$ 0.01$&$ 0.43$&      $ 0.02$& $0.12$ \\
        $V$/$I$&$ 0.02$& $0.12$&     $ 0.00$& $1.0 $ \rule[-1.2ex]{0pt}{0pt} \\
       
       \hline
      
       \end{tabular}
       \end{table}
      
 \subsubsection{Doppler-shift induced horizontal signal fluctuations}
 
To eliminate signal fluctuations resulting from Doppler shifts, we fitted a Gaussian to the Stokes $I$ spectrum in each spatial pixel. From these fits we calculated a Doppler map, which revealed a mean line core shift of 30 m\AA, while the standard deviation is smaller than our spectral resolution. Therefore the line core signals are largely restricted to a single image, labeled with 30 m\AA \ in Fig. \ref{fig:images}. Thus we expect to see scattering polarization signals mainly in the image taken at this wavelength. In the continuum intensity image the dark intergranules and bright granules are distinguishable. Although we refer in this paper to dark and bright features in the continuum intensity as intergranules and granules, we emphasize that these terms are ambiguous due to projection effects off the solar disk center.

 The contribution of Doppler-shift-induced fluctuations is on the order of 0.03\% or less. We estimated this upper limit with a simple calculation using the Doppler map mentioned above. The maximum calculated Doppler shift is 23 m\AA \ from the mean line core shifts. The standard deviation is 16 m\AA. We initially assumed the same Gaussian shaped Stokes $Q$/$I$ spectral signal in each spatial pixel located at the mean line core position of the spatially resolved data set. Further, we assumed that the $Q$/$I$ amplitude in each spatial pixel equals 0.19\%, this is, the average scattering $Q$/$I$ signal from the center-to-limb measurement at $\mu=0.6$. A shift of this initial Gaussian by the maximum Doppler shift leads to a drop in the a signal of about 0.03\%. Additionally, the correlation between Doppler velocities and continuum intensity is an order of magnitude smaller than between Stokes $Q$/$I$ and continuum intensity, which also indicates that the anti-correlation between $Q$/$I$ and continuum intensity is not an artifact.

  \subsubsection{Cross-talk from Stokes $I$}
  \label{sec:ct_stokesi}
   
\begin{figure}
    	\centering
    	\includegraphics[width=\hsize]{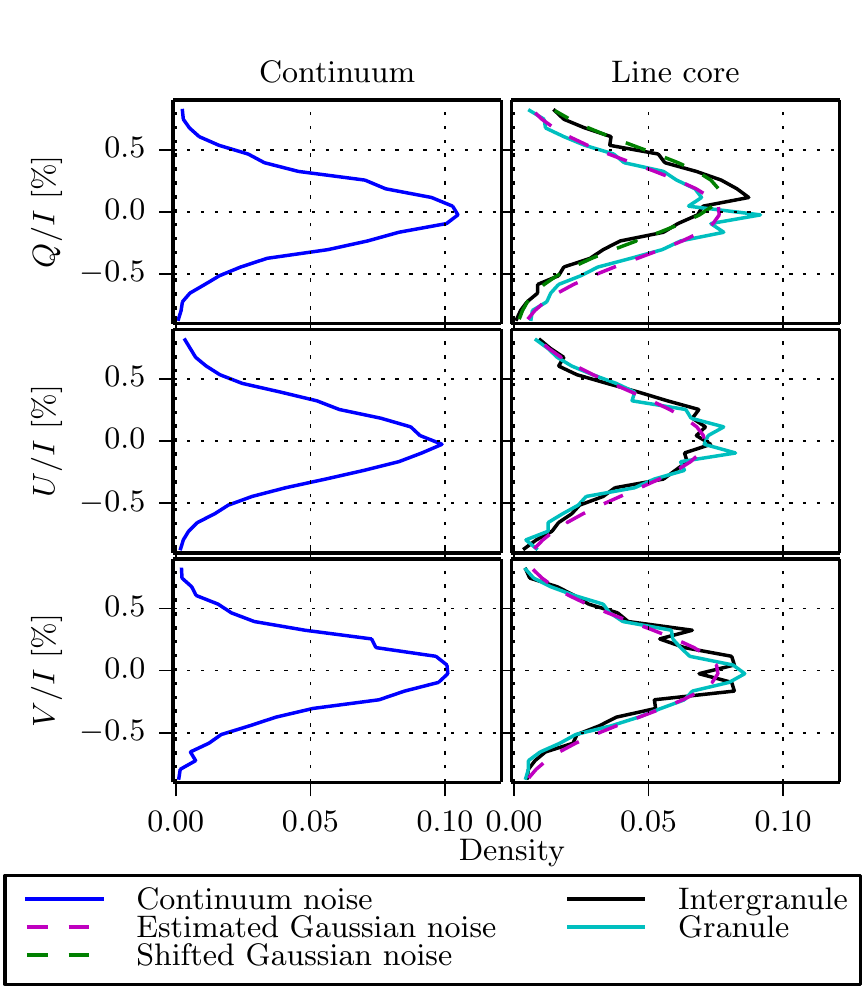}
    	\caption{
    		Left panels: probability density function of $Q$/$I$, $U$/$I$ and $V$/$I$ in the continuum. Right panels: noise distribution for the line core of Sr I inferred from that observed at the continuum wavelength in the various Stokes parameters (magenta dashed lines). The continuum image is classified into granules and intergranules, for which we plotted the Stokes signals separately. A shifted noise distribution (green dashed line) is shown to match the Sr I line core Stokes $Q$/$I$ signals in the intergranules.
    	}
    	\label{fig:noise}
    \end{figure}

We now describe a further test to show that although our Stokes signals are photon noise limited, we still have a clear distinction between Stokes $Q$/$I$ signals in granules and intergranules. This is an important test to rule out cross-talk from Stokes $I$ to Stokes $Q$/$I$ as a source of correlation between them. As no polarization is expected in the continuum, we fitted a Gaussian to the probability density function (see left panels of Fig. \ref{fig:noise}) to Stokes $Q$/$I$, $U$/$I$ and $V$/$I$ at the continuum wavelength. To estimate the photon noise expected in the Sr I line core, where less photons are available, the standard deviation of the fitted Gaussian from the continuum was rescaled. The rescaling procedure was done by calculating the parameters (i.e., the amplitude and standard deviation) of the line core Gaussian with two constraints: the first is the ratio of the line core intensity level to continuum intensity level of $0.5$ at $\mu=0.6$ (see right panel of Fig. \ref{fig:noise}) and the second is that the area below the Gaussian must be unity to get a probability density function. 

 Based on the continuum intensity image, we classified pixels to granule pixels and intergranule pixels. In order to emphasize the difference between them, we only analyzed a subset of all pixels. Thus granule pixels correspond to the 1000 brightest pixels and intergranule pixels correspond to the 1000 darkest pixels. Stokes signals corresponding to granules and intergranular lanes are plotted in the right panels of Fig. \ref{fig:noise}. The signals in $U$/$I$ and $V$/$I$ are noise dominated and the Gaussians for the intergranule and granule pixels are indistinguishable. But for the $Q$/$I$ signals in the interganule pixels, we needed to shift the center of the Gaussian in the right panel of Fig. \ref{fig:noise} from 0\% to 0.14\%.

  \subsubsection{Influence of spatial resolution}
To test if spatial resolution influences the sign of the line core $Q$/$I$ slope, we spatially smeared our data by convolving the line core intensity and polarization images with Gaussians with standard deviations between 0.5 pixel and 8 pixels and redid the correlation analysis. We find that the correlation decreases rapidly with the amount of smearing, but does not change its sign. The value of the slope of $Q$/$I$ versus continuum intensity in the line core when a Gaussian of a rms width of 8 pixels was applied (sampling of 0.64\arcsec/smeared pixel) is reduced to 0.76. This is close to the slope of $Q$/$I$ in the continuum at the original spatial resolution. Hence the difference in sign between our results and those of \citet{Malherbe2007} cannot be due to differences in spatial resolution.

  \subsubsection{Estimation of polarimetric structure size}
  
We estimated from our line core Stokes $Q$/$I$ image that the approximate diameter of structures with scattering polarization signals greater than the noise level is in the range of 0.5\arcsec \ to 1\arcsec. The estimation is based on a two-step analysis of the mean polarization signal, its standard deviation and the expected noise level in size-varying subregions in the image. We describe the steps in more detail in the next paragraph. \par
\begin{figure}
    	\centering
    	\includegraphics[width=\hsize]{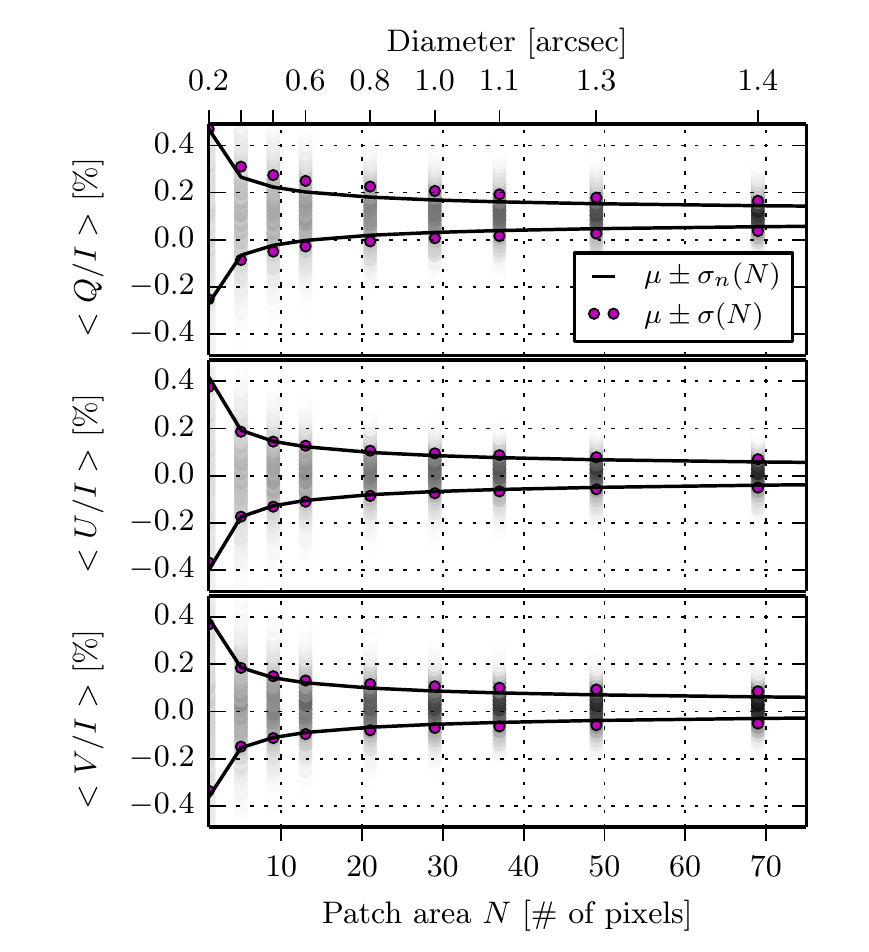}
  
     	\caption{
     		Histograms of the average polarization in all subregions with $N$ pixels. The histograms are represented by gray vertical bars, where black and white corresponds to 600 subregions and 0 subregions, respectively. Black lines: expected noise levels for $Q/I$, $U/I$ and $V/I$ when averaging over $N$ pixels; magenta dots: standard deviation of the histogram for each subregion; see text for details.}
    	
    	\label{fig:diameter}
    \end{figure}

In the first step, we defined a subregion in the image as a circular area, having a radius $R$ and containing $N$ pixels, centered on a given pixel in the image. As all pixels were used once as a central pixel, we had in total 6105 subregions per image, independently of the subregion size $N$. A pixel belonged to a given subregion if its Euclidean distance from the central pixel was smaller than $R$. If a central pixel was closer to the image boundary than $R$, the image was periodically continued. We then varied $R$ ($\propto \sqrt{N}$) and computed the mean polarization signal for each subregion. For each Stokes image and given $N$ we determined the distribution of the number of subregions as a function of average polarization signal using a histogram with bin size of 0.02\%. In Fig. \ref{fig:diameter} these histograms are depicted as vertical bars with a gray color scale. Black corresponds to 600 subregions and white corresponds to 0 subregions displaying a particular Stokes signal. A Gaussian was fitted to each histogram with the mean $\mu$ and the standard deviation $\sigma$ as free parameters. We plot $\mu \pm \sigma(N)$ to each histogram. Additionally, we calculated the noise level based on the single pixel noise $\sigma_n(N=1)$ present in each Stokes image (see Fig. \ref{fig:noise}). The noise level was reduced according to the number of averaged pixels assuming a shot noise process. According to this assumption, the noise $\sigma_n(N)=\sigma_n(1)/\sqrt{N}$ dropped with the square root of the number of pixels $N$ in the subregion. The noise level $\mu \pm \sigma_n(N)$ is indicated in Fig. \ref{fig:diameter} as well. 

  For $N=1$ the histograms are comparable to the right panel of Fig. \ref{fig:noise}  and are in agreement with the noise level, as expected. A minor offset in the mean of 0.01\% in the $U$/$I$ and 0.02\% in the $V$/$I$ image may be due to residual cross-talk from instrument polarization as explained in the earlier data reduction Sect. \ref{sec:dare_spr}. 
   
\begin{figure}
        	\centering
        	\includegraphics[width=\hsize]{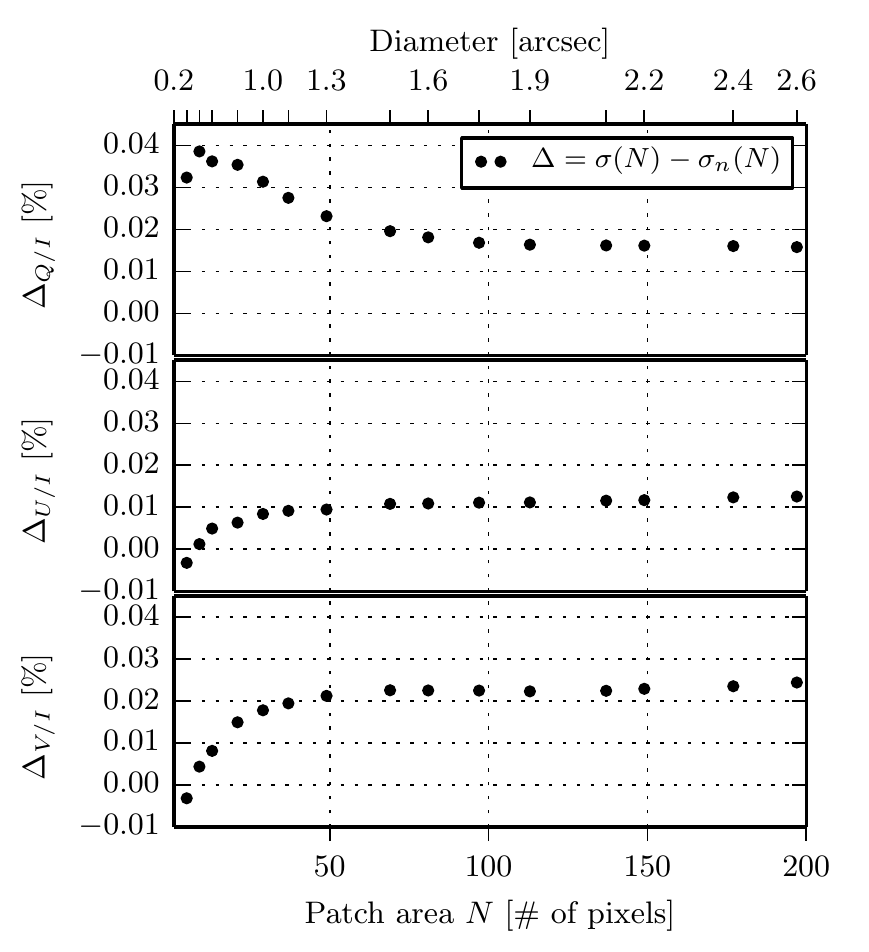}

         	\caption{
         		Absolute differences between the standard deviation $\sigma(N)$ obtained from the histograms and the expected noise $\sigma_n(N)$ for all three polarization Stokes parameters as a function of number of pixels per subregion $N$ (lower axis) and the subregion diameter (upper axis).}
        	
        	\label{fig:size} 
        \end{figure}

   In a second step we considered the absolute differences of the standard deviations of the histograms from the expected noise level. In Fig. \ref{fig:size} we plot the absolute difference $\Delta$ between the standard deviation $\sigma(N)$ and the expected noise $\sigma_n(N)$. The difference for $Q$/$I$ follows a decreasing trend for subregions with a diameter greater than 0.8\arcsec. This means that for subregion areas greater than 0.8\arcsec$^2$ only noise is added. The opposite occurs for $\Delta_{U/I}$ and $\Delta_{V/I}$. The differences increase for subregions with up to  $N$=40 pixels. This increase is likely due to small Zeeman signatures present in the data. The non-zero $\Delta$ for greater $N$ hints toward Zeeman signatures at larger scales combined with residual cross-talk errors.

\section{Discussion and conclusions}
We performed scattering polarization measurements in the Sr I 4607.3 \AA \ line with the Fast Solar Polarimeter mounted on TESOS at the VTT on Tenerife. To our knowledge, this is the first time a measurement of the scattering polarization in the Sr I line was done with a filtergraph instrument. 
In a first step, we have tested the reliability of the FSP by comparing the spatially averaged center-to-limb variation of the polarization signal in Sr I with previous results obtained with other instruments. In addition we verified that the center-to-limb variation of Stokes $Q$/$I$ in the neighboring scattering-insensitive Fe I 4607.6 \AA \ line is indeed consistent with zero signal.    
In a second step we analyzed a quiet Sun dataset in the Sr I line recorded at $\mu=0.6$ with a spatial resolution sufficient to clearly separate granules and intergranules in the Stokes $I$ continuum image. After temporal and spatial binning to a nominal resolution of 0.16\arcsec pixel$^{-1}$ and 2.5 s integration time per wavelength and Stokes parameter, the noise per pixel in the linear polarization is 0.3\%. We find a negative correlation of $r=-0.17$ between Stokes $Q$/$I$ in the line core and the Stokes $I$ continuum intensity. We rule out Doppler-shift induced fluctuations as the source of this correlation. We stress that our measurement is photon noise limited and seeing-induced cross-talk can be ruled out. Despite the small value of the correlation coefficient, our analysis shows statistically robust evidence of an anti-correlation of Stokes $Q$/$I$ with the continuum intensity in the Sr I line core.

The negative correlation found here seemingly contradicts the positive correlation reported by \citet{Malherbe2007} and \citet{Bianda2018}, both obtained by using a spectrograph and at $\mu=0.3$, hence, closer to the solar limb. The correlation coefficient by \citet{Bianda2018} is $r=0.19$, which is close to what has been found in this study, but with opposite sign. The spatial resolution achieved by \citet{Malherbe2007} cannot be better than the used slit width of 0.6\arcsec. We speculate that the difference in sign of the correlation found by these authors compared to that observed here could be due to the lower $\mu$ value at which the measurements were made. The contrast and shape of granulation and the sampled atmospheric height range changes considerably with $\mu$. This could lead to different signs of the correlation between Stokes $Q$/$I$ and continuum intensity. Results by \cite{DelPinoAleman2018} suggest that the positive sign is an artifact due to the reduced statistical significance arising from using a spectrograph.

In order to uncover the cause of different dependences of Stokes $Q$/$I$ in the line core of Sr I on continuum brightness at different $\mu$ values, both, lower noise, high resolution full Stokes measurements and theoretical calculations of the spatial structure of $Q$/$I$ in the Sr I line in realistic MHD simulations \citep{Rempel2014,Vogler2007,DelPinoAleman2018} at multiple $\mu$ values, are needed. From our measurements we furthermore conclude that in order to spatially resolve polarimetric signals the noise level needs to be significantly lower than $3\cdot 10^{-3}$.
 
 The closest $\mu$ value to our observations at which $Q$/$I$ images are presented in the most detailed published theoretical study of Sr I scattering polarization in 3D HD simulations is $\mu=0.5$  \citep{TrujilloBueno2007}. This polarization map with infinite spectral and high spatial resolution contains patches of polarization as large as 1\%, with most of the signals varying between 0\% and 0.7\%, in the case where horizontally fluctuating microturbulent magnetic fields in the Hanle saturation regime of $B=300$ G in the downflowing intergranules and $B=15$ G in granules at all heights, are considered. In the unmagnetized case, but with a degraded spectral resolution of 25 m\AA \ and spatial resolution corresponding to a 1 m telescope, the signals vary between $-0.08$\% and 0.9\%.
Patches of granular size with the above mentioned signal levels should be detectable in our data, but are not directly seen. 
With a statistical approach we estimated the sizes of the scattering polarization structures to be in the range between about 0.5\arcsec to 1\arcsec. The obtained size is on the order of the patch sizes of about 1\arcsec shown by \citet{TrujilloBueno2007}. 
However, a more definite conclusion requires that the simulations be degraded to the same spatial resolution and scattered light conditions as the observations.
Nevertheless, finding larger polarization signals in the intergranules is in qualitative agreement with \citet{TrujilloBueno2007} and results from a recently submitted paper by \cite{DelPinoAleman2018}. They state that scattering polarization emerges from local fluctuations of the radiation field  by plasma inhomogeneities, leading to axial symmetry breaking, even when observed at $\mu=1$. Polarization is therefore predominantly found at the interface of granules and intergranules, with a tendency for stronger polarization in intergranular lanes. Scatter plots provided by \cite{DelPinoAleman2018} show a negative correlation, independent of the $\mu$ value under consideration.

To conclude, in this work we have shown that instruments like FSP are now starting to reach the required combination of polarimetric sensitivity and spatio-temporal resolution to provide first observational feedback to theoretical studies of scattering polarization on small spatial scales. We have presented full Stokes filtergraph observations at $\mu=0.6$ in the Sr I 4607.3 \AA \ line and analyzed the images statistically, as the signal-to-noise ratio was not sufficient to directly detect local fluctuations in the solar scattering polarization signals at spatial scales significantly below 1\arcsec. From this analysis we found an anti-correlation between the line core Stokes $Q$/$I$ signals and the continuum intensity. We compared our findings with published observations, which were carried out at $\mu=0.3$ and showed a positive correlation. We additionally compared our result with 3D HD and MHD simulations, where the latter showed a negative correlation independent of the $\mu$ value.

   \begin{acknowledgements}
 The authors wish to acknowledge fruitful discussions with Rafael Manso Sainz and Ivan Milic. The Fast Solar Polarimeter project is funded by the Max Planck Society (MPG) and by the European Commission, grant No. 312495 (SOLARNET). We also would like to thank the staff of the Kiepenheuer Institut für Sonnenphysik in Freiburg for supporting our FSP observing campaigns at the VTT. We thank all the technical contributors not listed as co-authors for their invaluable input to the project. This project has received funding from the European Research Council (ERC) under the European Union's Horizon 2020 research and innovation program (grant agreement No. 695075) and has been supported by the BK21 plus program through the National Research Foundation (NRF) funded by the Ministry of Education of Korea. The participation of F. Zeuner was funded by the International Max Planck Research School for Solar System Science.
   \end{acknowledgements}
   
%

\begin{thebibliography}{28}
\expandafter\ifx\csname natexlab\endcsname\relax\def\natexlab#1{#1}\fi

\bibitem[{Beck {et~al.}(2010)Beck, Rubio, Kentischer, Tritschler, \& {del Toro
  Iniesta}}]{Beck2010b}
Beck, C., Rubio, L. R.~B., Kentischer, T.~J., Tritschler, A., \& {del Toro
  Iniesta}, J.~C. 2010, A{\&}A, 115, 10

\bibitem[{Bevington \& Robinson(2003)}]{Bevington2003}
Bevington, P.~R. \& Robinson, D.~K. 2003, {Data reduction and error analysis
  for the physical sciences}, 3rd edn. (McGraw-Hill)

\bibitem[{Bianda {et~al.}(2018)Bianda, Berdyugina, Gisler, Ramelli, Belluzzi,
  Carlin, Stenflo, \& Berkefeld}]{Bianda2018}
Bianda, M., Berdyugina, S., Gisler, D., {et~al.} 2018, A{\&}A, 614

\bibitem[{Borrero {et~al.}(2015)Borrero, Jafarzadeh, Sch{\"{u}}ssler, \&
  Solanki}]{Borrero2015}
Borrero, J.~M., Jafarzadeh, S., Sch{\"{u}}ssler, M., \& Solanki, S.~K. 2015,
  Space Sci. Rev., 1

\bibitem[{{De Wijn} {et~al.}(2009){De Wijn}, Stenflo, Solanki, \&
  Tsuneta}]{deWijn2009}
{De Wijn}, A.~G., Stenflo, J.~O., Solanki, S.~K., \& Tsuneta, S. 2009, Space
  Sci. Rev., 144, 275

\bibitem[{{Del Pino Alem{\'{a}}n} {et~al.}(2018){Del Pino Alem{\'{a}}n},
  {Trujillo Bueno}, St{\v{e}}p{\'{a}}n, \& Shchukina}]{DelPinoAleman2018}
{Del Pino Alem{\'{a}}n}, T., {Trujillo Bueno}, J., St{\v{e}}p{\'{a}}n, J., \&
  Shchukina, N. 2018, Astrophys. J., sub. [\eprint[arXiv]{1806.07293v1}]

\bibitem[{{del Toro Iniesta}(2004)}]{DelToroIniesta2004}
{del Toro Iniesta}, J.~C. 2004, {Introduction to Spectropolarimetry} No.~9
  (Cambridge University Press)

\bibitem[{Faurobert {et~al.}(2001)Faurobert, Arnaud, Vigneau, \&
  Frisch}]{Faurobert2001}
Faurobert, M., Arnaud, J., Vigneau, J., \& Frisch, H. 2001, A{\&}A, 378, 627

\bibitem[{Gandorfer(2002)}]{Gandorfer2002}
Gandorfer, A. 2002, {The Second Solar Spectrum: A high spectral resolution
  polarimetric survey of scattering polarization at the solar limb in graphical
  representation. Volume II: 3910 {\AA} to 4630 {\AA}} (Zurich: VdF)

\bibitem[{Iglesias {et~al.}(2015)Iglesias, Feller, \& Nagaraju}]{Iglesias2015}
Iglesias, F.~A., Feller, A., \& Nagaraju, K. 2015, Appl. Opt., 54, 5970

\bibitem[{Iglesias {et~al.}(2016)Iglesias, Feller, Nagaraju, \&
  Solanki}]{Iglesias2016}
Iglesias, F.~A., Feller, A., Nagaraju, K., \& Solanki, S.~K. 2016, A{\&}A, 590,
  A89

\bibitem[{Kentischer {et~al.}(1998)Kentischer, Schmidt, Sigwarth, \&
  Uexk{\"{u}}ll}]{Kentischer1998}
Kentischer, T.~J., Schmidt, W., Sigwarth, M., \& Uexk{\"{u}}ll, M.~v. 1998,
  A{\&}A, 340, 569

\bibitem[{Lagg {et~al.}(2016)Lagg, Solanki, {Vera Collados}, Franz, Feller,
  Kuckein, Schmidt, Ramos, Yabar, Denker, Balthasar, Volkmer, Staude, Hofmann,
  Strassmeier, Kneer, Waldmann, Borrero, Sobotka, Verma, Louis, Rezaei, Soltau,
  Berkefeld, Sigwarth, Schmidt, Kiess, \& Nicklas}]{Lagg2016a}
Lagg, A., Solanki, S.~K., {Vera Collados}, M., {et~al.} 2016, A{\&}A, 1

\bibitem[{{Landi Degl'Innocenti} \& Landolfi(2004)}]{LandiDeglInnocenti}
{Landi Degl'Innocenti}, E. \& Landolfi, M. 2004, {Polarization in spectral
  lines} (Kluwer Academic Publishers, Dordrecht)

\bibitem[{Malherbe {et~al.}(2007)Malherbe, Moity, Arnaud, \&
  Roudier}]{Malherbe2007}
Malherbe, J.-M., Moity, J., Arnaud, J., \& Roudier, T. 2007, A{\&}A, 462, 753

\bibitem[{Neckel(1999)}]{Neckel1999}
Neckel, H. 1999, Sol. Phys., 184, 421

\bibitem[{Rempel(2014)}]{Rempel2014}
Rempel, M. 2014, Astrophys. J., 789, 132

\bibitem[{Schou {et~al.}(2012)Schou, Borrero, Norton, Tomczyk, Elmore, \&
  Card}]{Schou2012}
Schou, J., Borrero, J.~M., Norton, A.~A., {et~al.} 2012, Sol. Phys., 275, 327

\bibitem[{Steiner \& Rezaei(2012)}]{Steiner2012}
Steiner, O. \& Rezaei, R. 2012, ASP Conf. Ser., 456

\bibitem[{Stenflo(1994)}]{Stenflo1994}
Stenflo, J.~O. 1994, {Solar Magnetic Fields}, Vol. 189 (Springer
  Science+Business Media, Dordrecht), 401

\bibitem[{Stenflo(2011)}]{Stenflo2011}
Stenflo, J.~O. 2011, A{\&}A, 529, A42

\bibitem[{Stenflo(2014)}]{fts}
Stenflo, J.~O. 2014, {FTS Atlas 2:
  http://data.irsol.ch/data{\_}archive/{\#}ftsv}

\bibitem[{Stenflo {et~al.}(1997)Stenflo, Bianda, Keller, \&
  Solanki}]{Stenflo1997a}
Stenflo, J.~O., Bianda, M., Keller, C.~U., \& Solanki, S.~K. 1997, A{\&}A, 322,
  985

\bibitem[{Stenflo \& Keller(1997)}]{Stenflo1997}
Stenflo, J.~O. \& Keller, C.~U. 1997, A{\&}A, 934, 927

\bibitem[{Tritschler {et~al.}(2002)Tritschler, Schmidt, Langhans, \&
  Kentischer}]{Tritschler2002}
Tritschler, A., Schmidt, W., Langhans, K., \& Kentischer, T. 2002, Sol. Phys.,
  211, 17

\bibitem[{{Trujillo Bueno} \& Shchukina(2007)}]{TrujilloBueno2007}
{Trujillo Bueno}, J. \& Shchukina, N. 2007, Astrophys. J., 664, L135

\bibitem[{{Trujillo Bueno} {et~al.}(2004){Trujillo Bueno}, Shchukina, \&
  {Asensio Ramos}}]{Bueno2004}
{Trujillo Bueno}, J., Shchukina, N., \& {Asensio Ramos}, A. 2004, Nature, 430,
  326

\bibitem[{V{\"{o}}gler \& Sch{\"{u}}ssler(2007)}]{Vogler2007}
V{\"{o}}gler, A. \& Sch{\"{u}}ssler, M. 2007, A{\&}A, 465, L43

\end{thebibliography}
%

\end{document}